\documentclass[10pt, conference]{IEEEtran}
\IEEEoverridecommandlockouts
\usepackage{cite}
\usepackage{amsmath,amssymb,amsfonts}
\usepackage{algorithmic}
\usepackage{graphicx}
\usepackage{textcomp}
\usepackage[dvipsnames]{xcolor}
\def\BibTeX{{\rm B\kern-.05em{\sc i\kern-.025em b}\kern-.08em
    T\kern-.1667em\lower.7ex\hbox{E}\kern-.125emX}}

\usepackage{subcaption}

\usepackage{tabularx}
\usepackage{multirow}
\usepackage{booktabs}
\usepackage{tablefootnote}
\usepackage{url}

\usepackage{textcase}

\begin{document}

\title{A Data Set of Generalizable Python Code Change Patterns
}

\author{\IEEEauthorblockN{Akalanka Galappaththi}
\IEEEauthorblockA{
\textit{University of Alberta, Edmonton, Canada}\\
	akalanka@ualberta.ca}
\and
\IEEEauthorblockN{Sarah Nadi}
	\IEEEauthorblockA{
	\textit{University of Alberta, Edmonton, Canada}\\
	nadi@ualberta.ca}

}

\maketitle

\begin{abstract}
Mining repetitive code changes from version control history is a common way of discovering unknown change patterns.
Such change patterns can be used in code recommender systems or automated program repair techniques.
While there are such tools and datasets exist for Java, there is little work on finding and recommending such changes in Python.
In this paper, we present a data set of manually vetted generalizable Python repetitive code change patterns.
We create a coding guideline to identify generalizable change patterns that can be used in automated tooling.
We leverage the mined change patterns from recent work that mines repetitive changes in Python projects and use our coding guideline to manually review the patterns.
For each change, we also record a description of the change and why it is applied along with other characteristics such as the number of projects it occurs in.
This review process allows us to identify and share 72 Python change patterns that can be used to build and advance Python developer support tools.

\textit{Data set link: \url{https://figshare.com/articles/conference_contribution/Generalizable_Python_code_change_patterns/21966677}}
\end{abstract}


\section{Introduction}


Software systems evolve over time as developers change their code to implement new features, fix bugs, or find more concise, readable, or efficient ways to implement the functionality.
Repetitive source code changes that developers make can indicate useful improvements that can be used as a basis for code recommender systems or automatic program repair tools.
For example, by analyzing repetitive changes, we can observe that, in Java, developers often replace a loop that adds elements from a given collection to a list one at a time with the \texttt{addAll} method \cite{ChangePatNguyen2019}.
Such a change makes the code more concise and readable, and can improve efficiency in some cases.

There is a long line of research that leveraged this idea to mine frequent code change patterns from version-control history and then use these patterns to recommend bug fixes~\cite{GetaFixBader2019, BugFixPan2009, LaseMeng2013}, repair static analysis violations \cite{PhoenixYoshida2020, DevReplayUeda2022}, enhance code \cite{LaseMeng2013, RevizorSmirnov2021}, or migrate APIs \cite{APIMigrateAlrubaye2019}.
Most of this change pattern mining research focused on Java projects \cite{ChangePatNguyen2013, ChangePatNegara2014, TangledChangesHerzig2013, ChangePatNguyen2019,CDistFluri2007} with only one recent work on Python \cite{ChangePatDilhara2022} despite Python being the most used programming language \cite{TobeIndex}. 

\begin{figure}[t!]
	\includegraphics[scale=0.26]{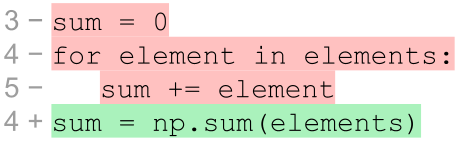}
	\caption{Example of a change pattern that replaces a \texttt{for} loop (before version) with an API (after version) \cite{ChangePatDilhara2022}.\vspace{-0.6cm}}
	\label{fig:high_level_trans}
\end{figure}

Inspired with the Java tooling above, we wanted to create a Python linter tool that can flag instances of frequently changed code to notify developers that there are better ways for implementation.
We thus looked at recent work by Dilhara et al.~\cite{ChangePatDilhara2022} that adapted Java frequent change-pattern mining tools to work for Python. 
With the goal of understanding Python code evolution as an educational resource for developers as well as to create a data set that can be used to provide a basis for code change recommendations and to build automated tooling, the authors analyzed histories of 1,000 Python machine-learning projects to mine code change patterns.
The authors considered a pattern \textit{meaningful} if the pattern can be described as a high-level program transformation. Figure \ref{fig:high_level_trans} shows an example of such a high-level program transformation where a \verb|for| loop that calculates the sum of elements is replaced with a call to the \verb|numpy.sum| Application Programming Interface (API) function. Through their analysis, they identified 561 meaningful code change patterns \cite{MLRepPat}.
Thus, we set out to use this data set as the basis for our linter tool. However, when investigating the change patterns more carefully, we found that some of the patterns do not represent generalizable changes that can be used for automated tools. 
In this paper, we consider a change pattern as \emph{generalizable} if the code after the change is semantically similar to the code before the change, and if the change is generally applicable. For example, when inspecting the source code changes in Figure \ref{fig:high_level_trans}, a developer can notice that the API call in line 4 of the after version is semantically similar to lines 3-5 of the before version. When another project has a similar before version structure (a loop that sums items of an iterator), it is possible to recommend replacing it with the corresponding after version. 

\begin{figure}[t!]
  \begin{subfigure}{0.49\textwidth}
    \includegraphics[scale=0.26]{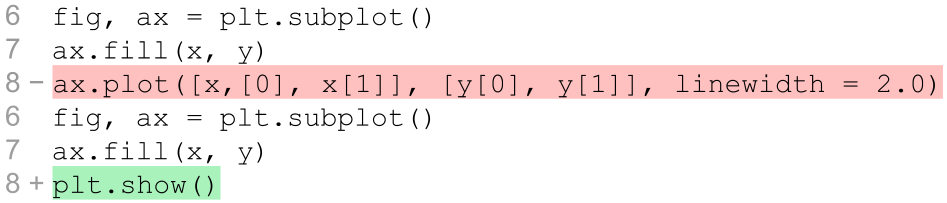}
	\caption{Replacing \texttt{plot} API with \texttt{show} API}
	\label{m_ex_a}
  \end{subfigure}%
  
  \begin{subfigure}{0.49\textwidth}
    \includegraphics[scale=0.2]{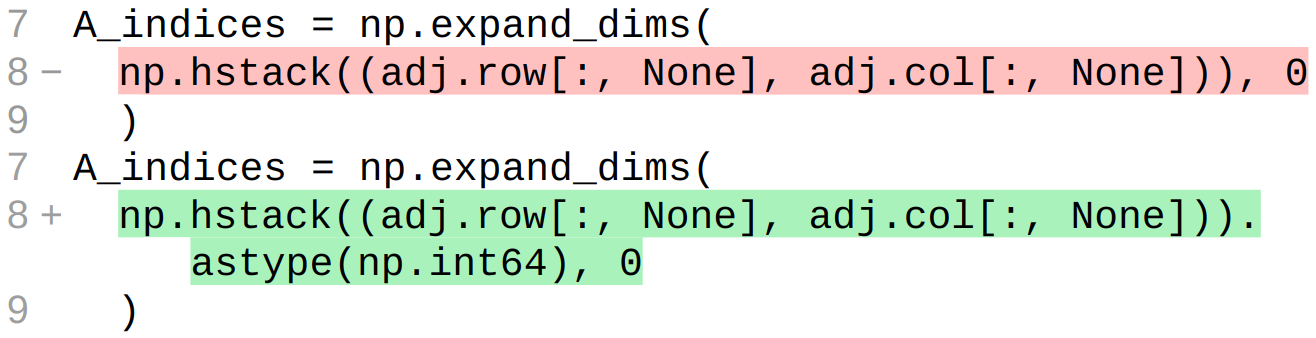}
    \vspace{-0.2cm} 
	\caption{Change the matrix type}
	\label{m_ex_b}	
  \end{subfigure}%

\caption{Change patterns that are not genaralizable  \cite{ChangePatDilhara2022}\vspace{-0.6cm}} \label{fig:meaningless}
\end{figure}

Figure \ref{fig:meaningless} shows two examples from Dilhara et al.~\cite{ChangePatDilhara2022}'s 561 code change patterns that are not generalizable, according to our definition.
In the change pattern in Figure \ref{m_ex_a}, \verb|ax.plot(...)| and \verb|plt.show()| are two different operations. 
While the mining algorithm may have found enough instances of such a change to mark it as a pattern, it does not make sense to flag any usage of \verb|ax.plot(...)| and ask developers to replace it with \verb|plt.show()| since the former creates the plot while the later shows it.
In other words, these are semantically dissimilar operations altogether.
Figure \ref{m_ex_b} shows another problematic change pattern. 
While the change involves similar operations with a different matrix type in the after version, this change pattern is not generally applicable, because changing to type \verb|np.int64| will not always apply. Another developer may need to use different types such as \verb|int32| or \verb|float32| depending on the data that they handle. 

The above examples suggest that while automatically mining code change patterns from project commit histories allows discovering previously unknown patterns, patterns returned by the mining algorithms are not always generalizable or correct.
This problem is not specific to Dilhara et al.~\cite{ChangePatDilhara2022}'s work, but is rather general to the concept of frequent change mining~\cite{ChangePatNguyen2019, ChangePatNguyen2013, AvatarLiu2019, MineChanges2014, APIRecNguyen2016}. 
This is because the mining algorithms investigate the syntax changes to find patterns but do not consider the semantics of the change. 

Therefore, to use change patterns as a basis for any automated tooling, it is important to find a way to automatically detect generalizable change patterns. 
Thus, the question becomes how can we identify genaralizable change patterns that can be used for recommendations? In this data paper, we follow a well-defined and documented manual validation approach to identify genaralizable patterns in the data set provided by Dilhara et al.~\cite{ChangePatDilhara2022}. By filtering out the patterns that are not genaralizable, we provide a data set that contains code transformations that can be generally applicable to Python projects. 
Further, we annotate the patterns with a description of the change and various quantitative characteristics such as number of developers who apply a change, number of projects where the pattern was found, and invocations of any third-party APIs. Based on this, we contribute a data set of 72 verified Python change patterns that can be used to make code change recommendations or to derive automated criteria to identify genaralizable code change patterns \cite{artifact}.

\section{Data Set Construction}
\label{sec:methods}
In this section, we provide a brief overview of the existing data set that we leverage in our work and the evaluation steps we conduct to identify generalizable change patterns.

\subsection{Original Python Code change patterns~\cite{ChangePatDilhara2022}}

Dilhara et al.~\cite{ChangePatDilhara2022} adapted \textsc{CPATMiner}~\cite{ChangePatNguyen2019} to find code change patterns in Python projects. The authors run the tool on commits of 1,000 top-rated Python machine learning applications from diverse application domains. The automatic mining process resulted in over 28,000 code change patterns. Each pattern has at least three change instances, because the authors set the frequency threshold to three to consider a repetitive code change as a pattern. 
To evaluate the usefulness and nature of their mined patterns, they manually evaluate a selected sample by picking the top 500 patterns across five different dimensions: frequency, size, frequency $\times$ size, number of projects where the pattern was found, and number of developers who made the change. 
They consider a change pattern as \textit{meaningful} if it can be described using a high-level program transformation.
While there is no explicit coding guide provided, we interpret ``high-level program transformation'' to mean being able to describe the change as \emph{transform X to Y}, \emph{replace X with Y}, or \emph{move X into Y}.
Based on this criteria, the authors selected 561 patterns out of 2,500 patterns that they inspected. 
They then categorize the 561 change patterns into pattern groups and the pattern groups into trends. The first three columns of Table \ref{tab:meaningful_patterns} provide an overview of the trends, pattern groups, and the number of patterns in each group.
As illustrated in the introduction, the ability to describe a code change as a high-level program transformation~\cite{ChangePatDilhara2022} is not enough to exclude useless change patterns.

\subsection{Identifying Meaningful Change Patterns}
For our systematic manual investigation of generalizable code change patterns, we consider the 561 code change patterns that were already vetted by the original authors.
Since these 561 change patterns were sampled across five different dimensions, they provide a subset of more likely stable code change patterns \cite{ChangePatNguyen2013}, as opposed to randomly sampling from the full 28,000 patterns.

To reduce the number of change patterns we manually investigate, we first filter out patterns where all change instances were made by the same developer. This is because a pattern is more likely to be applicable and generalizable if multiple developers felt the need to make the change. 
Based on this criteria, we discard 134 patterns and proceed with 427 patterns.

\begin{figure}[t!]
  \begin{subfigure}{0.49\textwidth}
  \centering
    \includegraphics[scale=0.26]{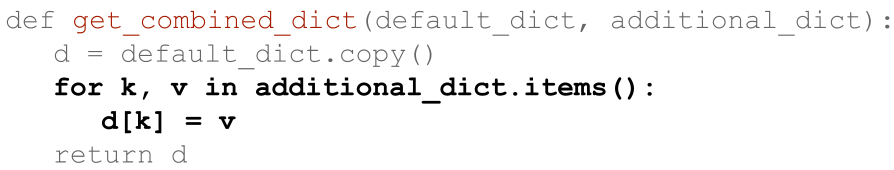}
	\caption{Before version that uses a \texttt{for} loop to update a dictionary}
	\label{p_a}
  \end{subfigure}%
  
  \begin{subfigure}{0.49\textwidth}
  \centering
    \includegraphics[scale=0.26]{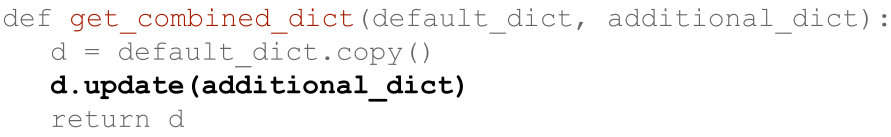}
	\caption{After version that uses \texttt{update} API to update a dictionary}
	\label{p_b}	
  \end{subfigure}%

\caption{Change pattern that dissolves a \texttt{for} loop with a built-in Python API \cite{ChangePatDilhara2022} \vspace{-0.6cm}} \label{fig:pattern}
\end{figure}

We now move on to manually review the remaining 427 change patterns.
To guide this review process, we need to create a well-defined coding guideline with clear criteria for identifying generalizable patterns.
To create this guide, we randomly select three pattern groups (\emph{Transform to function in list or dict}, \emph{Swap data visualization}, and \emph{Update container}) to identify differentiating characteristics of patterns. Each pattern contains a sample code snippet with before and after versions with emphasis on the change pattern captured by the pattern mining algorithm. Figure \ref{fig:pattern} shows a pattern that dissolves a \verb|for| loop that updates the values of a Python dictionary, replacing it with the built-in Python API, \verb|update|. Darker text in Figure \ref{p_a} and Figure \ref{p_b} indicate the changes that are part of the change pattern, as captured by the mining algorithm. 

We develop the following coding guide for identifying generalizable change patterns.

\begin{LaTeXdescription}

  	\item[1.] \textbf{Semantic similarity of the before and after change:} Check if the changed parts of the before and after version of the code reflect semantically similar operations. If \emph{No}, discard the pattern.
	\item[2.] \textbf{General applicability of the pattern:} If the answer to Criterion 1 is \emph{Yes}, check for the general applicability of the pattern. If the pattern contains attributes or functions that specifically belong to the project or the change is valid only in specific circumstances, discard the pattern.
	
	
  
\end{LaTeXdescription}

\begin{figure}[t!]
	\includegraphics[scale=0.26]{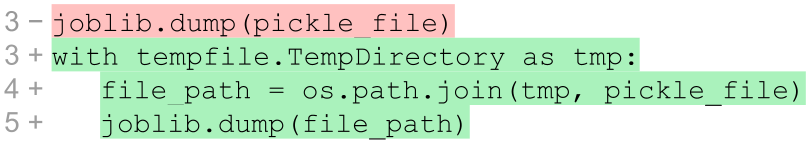}
	\caption{Example non-generalizable pattern using a context manager with a temp directory to dump a \emph{pickle} file  \cite{ChangePatDilhara2022}. \vspace{-0.9cm} }
	\label{fig:context}
\end{figure}

We use Criteria 1 and 2 of the coding guide to filter out code change patterns like Figures \ref{m_ex_a} and \ref{m_ex_b}, respectively.
Additionally, Criterion 2 of the coding guide filters out the patterns if the after version of the pattern is only valid for a specific situation. For example, in Figure \ref{fig:context}, we cannot always recommend developers to use \verb|tempfile.TempDirectory| context manager when dumping a \emph{pickle} file.
Overall, our coding guideline ensures that the code change patterns that we mark as generalizable are possible to use as code change recommendations for developers.

To make our data set more useful for use in code change recommender systems or for identifying criteria that can be used to automatically identify generalizable patterns, we also find and record additional characteristics of the generalizable change patterns we identify.
We include information related to external libraries and Python modules used in each code change pattern such that the data can be filtered to make recommendations for specific application domains. For example, generalizable code change patterns that use PyTorch or TensorFlow can be used to make code change recommendations that are specific to machine learning applications. Additionally, we record the number of developers and number of projects of
each code change pattern. Finally, we make notes for patterns that filtered out in the annotation process.

After the annotation process, we investigate the patterns we marked as generalizable to understand why the change was made and why such a change might be useful to recommend. We investigate library/Python documentation, online Q\&A forums, and commit messages of the change instances that belong to each pattern to collect information about the change. 

\section{Data Set of Generalizable Code Change Patterns}

%
%


\paragraph*{Num. of generalizable patterns} We find that in 260 (61\%) patterns, the changed code does not involve semantically similar operations. From the remaining patterns, we find that 95 (22\%) patterns are not applicable as general patterns (discarded by Criteria 2 in the coding guide). 
This leaves us with 72 (17\%) generalizable code change patterns that are useful to make recommendations. 

\begin{table*}[t!]
    \centering
    \caption{Number of generalizable patterns in each pattern group. Columns 1-3 are taken from Dilhara et al.~\cite{ChangePatDilhara2022} while the last two columns are the number of generalizable patterns and example reasons we identify through our review process.}
        \resizebox{0.85\textwidth}{!}{
	\begin{tabular}{r l l r r l}

		\toprule
		
		\multicolumn{1}{c}{\textbf{\#}} &
		\multicolumn{1}{l}{\textbf{Trend}} & 
		\multicolumn{1}{l}{\textbf{Pattern group}} &
		\multicolumn{1}{c}{\textbf{\# of}} &
		\multicolumn{1}{c}{\textbf{\# of generalizable}} &
		\multicolumn{1}{l}{\textbf{Example reason}} \\
		
		& & &
		\multicolumn{1}{c}{\textbf{patterns}} &
		\multicolumn{1}{c}{\textbf{patterns (\%)}} &
		\\
		
		\midrule
		
		1 & \multirow{8}{0.12\linewidth}{Move to \texttt{with} statement and use context manager}
		& Read, write, traverse data & 135 & 31 (23\%) & avoid mishandling resources\\
		2 & & Disable or enable gradient calculation & 27 & 0 (0\%) & \\
		3 & & Swap ML training device & 5 & 0 (0\%) & \\
		4 & & Change name and variable scope in DL networks & 30 & 0 (0\%) & \\
		5 & & Execute dependencies of a TansorFlow graphs & 16 & 0 (0\%) & avoid mishandling dependencies\\
		6 & & Temporarily change configuration of libraries & 7 & 1 (14\%) & avoid mishandling error states\\
		7 & & Move to context manager in pytest & 85 & 6 (7\%) & avoid mishandling raised exceptions\\
		8 & & Use context manager to open temorary directories & 28 & 2 (7\%) & avoid mishandling temporary files\\
		
		\midrule
		
		9 & \multirow{7}{0.12\linewidth}{Dissolve \texttt{for} loops into domain specific abstractions} 
		& Transform to optimized operations in Numpy & 51 & 8 (16\%) & run time efficiency\\
		10 & & Transfrom to functions in \emph{list} or \emph{dict} & 7 & 2 (29\%) & run time efficiency\\
		11 & & Transform to Python built-in functions & 5 & 0 (0\%) & \\
		12 & & Transform to function in \emph{String} & 4 & 1 (25\%) & run time efficiency\\
		13 & & Transform to \emph{set} operation & 2 & 1 (50\%) & run time efficiency\\
		14 & & Use optimized operations in PyTorch & 4 & 0 (0\%) & run time efficiency\\
		15 & & Use optimized operations in TensorFlow & 4 & 1 (25\%) & run time efficiency\\
		                   
		\midrule
	
		16 & \multirow{6}{0.12\linewidth}{Update API usage} 
		& Migrate to APIs in ML libraries & 3 & 1 (33\%) & run time efficiency \\
		17 & & Transform matrix & 13 & 0 (0\%) & \\
		18 & & Swap data visualization & 4 & 0 (0\%) & \\
		19 & & Composite ML APIs & 3 & 1 (33\%) & deprecated API\\
		20 & & Update container & 15 & 2 (13\%) & access items of an iterator\\
		21 && Update type of matrices & 4 & 0 (0\%) & \\
		
		\midrule
		
		22 & \multirow{3}{0.12\linewidth}{Use advanced language features}
		& Simplify conditional statement & 3 & 1 (33\%) & improve readability \\
		23 & & Migrate from \emph{Dict}, \emph{Set}, \emph{List} constructors to literals & 4 & 4 (100\%) & run time efficiency\\
		24 & & Transform to Python \emph{List}, \emph{Dict}, \emph{Set} comprehension & 101 & 10 (10\%) & improve readability/Pythonic way\\
		
		\midrule

		\multicolumn{3}{l}{Total} & 561 & 72 (17\%) & \\		
		
		\bottomrule	
	\end{tabular}
	}\vspace{-0.5cm}
	\label{tab:meaningful_patterns}
	
\end{table*}

The fourth column in Table \ref{tab:meaningful_patterns} shows the number and proportion of generalizable patterns by pattern group. We find that in 9 pattern groups, 20\%-100\% of the contained change patterns are generalizable, while the remaining 15 pattern groups have no generalizable patterns or a very low proportion.

\paragraph*{Reasons for change} For each generalizable pattern we identify, we include a free-text description and reason for the change, which we summarize in the last column of Table~\ref{tab:meaningful_patterns}. 
Pattern groups 1-8 deal with context managers~\cite{PythonDoc:with}. 
Pattern group 1 deals with changing code to use a context manager so that developers do not need to worry about accidentally forgetting to close files. Similarly, pattern groups 6-8 use context managers to include dependencies, handle error states, handle raised errors, and create temporary files respectively.

The second trend (pattern groups 9 - 15, except 11) all focus on run-time efficiency.
Using APIs in Python and third-party libraries like \emph{Numpy} is generally more efficient compared to rewriting the same logic using loops \cite{NumpyWalt2011}. For example,  pattern group 9 use APIs like \verb|indices|, \verb|mean|, and \verb|einsum| to dissolve \verb|for| loops, resulting in performance improvement. 
Similarly, the generalizable pattern from group 12 uses \verb|str.join| to concatenate strings instead of appending inside a loop.

The third trend focuses on updating API usages in general, but we find that the change in the generalizable pattern we identify from group 16 improves performance (using \emph{Numpy.average} instead of calculating the average by dividing the sum by size). In contrast, the generalizable pattern from group 19 \emph{composite ML APIs} uses a new API to replace a deprecated API. One of the two patterns in group 20, \emph{update container}, uses a \verb|list| container to retrieve the elements in a reversed list. 
The other generalizable pattern from group 20 does not change the container but swaps the \verb|dict| constructor with \verb|dict| literal. We believe this pattern is better suited under \emph{migrating from Dict, Set, List constructors to literals}. 

The generalizable change pattern from group 22 reduces the length of a conditional statement by using the \verb|in| operator. This pattern focuses on improving readability by making the code concise. All the generalizable patterns from group 23 change constructors to literals to improve runtime \cite{IdiomsZhang2022}. This change eliminates additional searching for Python builtins like \texttt{list, dict, set} \cite{SO}. 
Pattern group 24 had many generalizable change patterns that use Python comprehensions, i.e., Python idioms that make the code more Pythonic \cite{IdiomsZhang2022}. They also improve the code readability \cite{IdiomsZhang2022}. 

\paragraph{Libraries, Devs, \& Projects} In terms of libraries, we find 9 of the confirmed generalizable patterns use external libraries \emph{Numpy}, \emph{TensorFlow}, \emph{Spacy}, \emph{h5py}, and \emph{mpmath}. 
We also find that 80\% of the generalizable patterns had 2 or 3 unique developers who perform the change. When looking at the number of projects where the instances of generalized patterns were found, 74\% of those patterns had 2 or 3 unique projects. Only seven (9\%) patterns had instances extracted from a single project.


\paragraph*{Data Format} We share the data set on our artifact page \cite{artifact-fig}. We provide a spreadsheet similar to the structure of Table \ref{tab:meaningful_patterns}. We include each generalizable pattern under its corresponding pattern group. Each pattern contains a free-text reason for the change, a link that directs to additional information such as the before and after version of the pattern, and project instances where the pattern was found (during the original mining process~\cite{ChangePatDilhara2022}). Additionally, we include the original data set that we annotated which includes the non-generalizable patterns as well.


\section{Discussion and Data Uses}

This paper presents a data set of manually verified generalizable Python code change patterns.
This data set is useful to provide Python code change recommendations, build tools that automatically apply code transformation, or to explore differentiating characteristics for automatically identifying generalizable change patterns. 
For example, Python refactoring tools that change non-idiomatic code to idioms~\cite{IdiomsZhang2022} can extend their work by leveraging the code change patterns specific to Python idioms in this data set (e.g., pattern group 24).

Although the data set we present is small compared to the original mined patterns~\cite{ChangePatDilhara2022}, the 72 patterns we provide are manually validated for their general applicability. 
Directly using the mined frequent change patterns without this vetting can result in meaningless or incorrect recommendations or transformations that frustrate developers.
Thus, this manual review work is an essential first step for building useful tooling.
Our data set saves other researchers the manual review time (35 hours) and paves the way forward for creating more Python developer support tools.
The resulting data set contains a diverse set of patterns that cover 15 different pattern groups. The patterns in the data set improve code readability, improve performance, provide a better way of handling resources like files, or inform about API migration for deprecated APIs. The reasons we record could be provided along with code change recommendations to strengthen developers' trust to accept recommendations.

Since our data contains both change patterns we mark as generalizable and those we discard, future research can investigate how to infer differentiating characteristics.
Finding characteristics that automatically identify change patterns suitable for automated support tools can help advance research efforts on mining repetitive changes~\cite{ZimmermanMining05,APIRecNguyen2016,LaseMeng2013,DevReplayUeda2022}.
\section{Conclusion}
We reviewed 561 existing change patterns that were mined from the histories of 1,000 Python projects~\cite{ChangePatDilhara2022} and contributes a data set of 72 Python code change patterns, along with the description and reason for each change.
Overall, we find that only 17\% of these patterns are generalizable enough to be used as a basis for code recommender systems or automated repair tools.
This emphasizes the need to vet mined change patterns before using them to build developer support tooling.
Our contributed data set~\cite{artifact-fig} thus paves the way for building more automated Python support tooling.

\section*{Acknowledgements}
We would like to thank Malinda Dilhara and Danny Dig for their feedback on this work.

\bibliographystyle{IEEEtran}  
\bibliography{references}

\end{document}